# SIMULATED LORENTZ FORCE DETUNING COMPENSATION WITH A DOUBLE LEVER TUNER ON A DRESSED ILC/1.3 GHZ CAVITY AT ROOM TEMPERATURE

C. Contreras-Martinez[†], Y. Pischalnikov, J.C. Yun, Fermilab, Batavia, Illinois, USA

*Abstract*

Pulsed SRF linacs with high accelerating gradients experience large frequency shifts caused by Lorentz force detuning (LFD). A piezoelectric actuator with a resonance control algorithm can maintain the cavity frequency at the nominal level, thus reducing the RF power. This study uses a double lever tuner with a piezoelectric actuator for compensation and another piezoelectric actuator to simulate the effects of the Lorentz force pulse. A double lever tuner has an advantage by increasing the stiffness of the cavity-tuner system, thus reducing the impact of LFD. The tests are conducted at room temperature and with a dressed 1.3 GHz 9-cell cavity.

## INTRODUCTION

The primary source of cavity detuning for linacs operated in high gradient and high beam loading is Lorentz force detuning (LFD). The interaction of the RF magnetic field in the cavity and the wall currents on the cavity gives rise to the Lorentz force. The interaction with the magnetic field caused the equator of the cavity to bend inwards while the magnetic field interaction causes the cavity to bend outwards. Sending more RF power to the cavity can maintain the nominal accelerating gradient, but this has limitations. Piezoelectric (piezos) actuators in conjunction with a resonance control algorithm are widely used to keep the accelerating gradient of the cavity constant. The piezos can expand or contract depending on the polarity of the applied voltage.

The test's purpose is to measure the dynamic parameters of the dressed cavity LCLS-II tuner system [1] and demonstrate that there is no significant difference from the previous test at S1Global, which showed successful compensation of LFD on all the types of cavity-tuner systems [2]. This tuner has high stiffness, and the design allows to replace the piezo and stepper motors through a designated port on the cryomodule. The piezo actuator is excited by a single cycle sine wave before the arrival of the RF pulse. The duration, time advance, and amplitude of the sine wave are optimized manually by trial and error. This method is standard and used by many groups [3].

## EXPERIMENTAL SETUP

A schematic of the hardware and signal topology for measurement of the cavity frequency detuning and resonance control is shown in Figure 1. An RF analog signal generator produces the input signal to excite the cavity in the $\pi$ mode at 1298.838 MHz at room temperature. The forward power is coupled through a directional coupler and fed to input A of the AD8032 Analog Phase Detector (APD). The transmitted power of the cavity is sent to input B of APD. The output signal of the APD is proportional to the phase shift between the forward and transmitted power of the cavity. The forward and transmitted power phase can then be related to the cavity detuning.

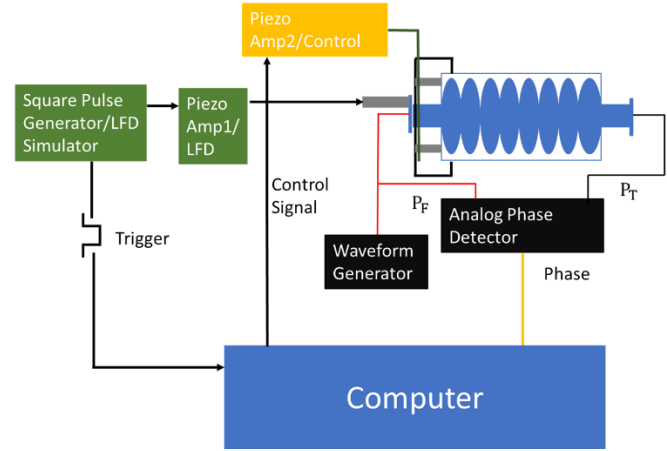

Figure 1: Schematic of signals used to measure the cavity frequency and for resonance control. A square wave pulse generator was used to simulate the LFD pulse and to trigger the control.

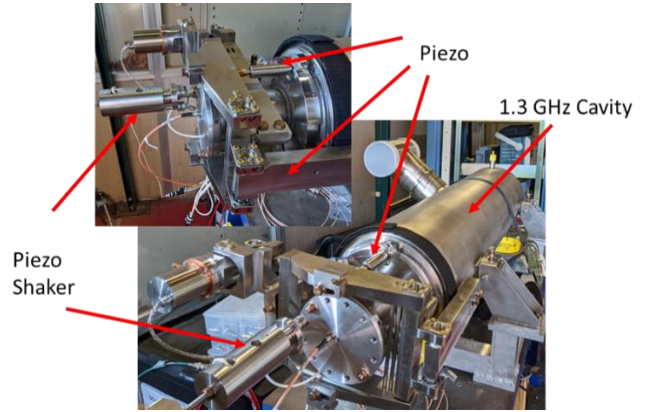

Figure 2: Setup of LFD simulation with 1.3 GHz cavity. Two piezo capsules are used for resonance control and a piezo (denoted as piezo shaker) is used to simulate the LFD.

The formula to relate the phase of the two signals to the cavity detuning is given by

$$\Delta f = \frac{f_0}{2Q_L} \tan \phi \quad (1)$$

where $\Delta f$ is the cavity detuning, $f_o/2Q_L$ is the half-bandwidth of the cavity, and $\phi$ is the phase between the forward and transmitted power. The cavity frequency detuning was digitized with NI-PXI-4472 14-bit ADC. The cavity detuning calculation, recording, and resonance control of the cavity detuning data are done through LabVIEW.

A square pulse with a small duty factor simulates the LFD pulse. The pulse is also used as a trigger for data acquisition and to excite the control signal. The piezos used to control and simulate the LFD pulse are on the same side, which contains a bellow (see Figure 2). The LFD pulse piezo will be called the shaker piezo and the other piezos will be called control piezos. The shaker piezos sit directly on the cavity flange while the control piezos connect with the cavity bellow via a split ring interface.

## LFD RESONANCE CONTROL

The simulated LFD pulse is done with a square wave pulse on the shaker piezo as shown in Figure 3. The pulse is not entirely square due to the capacitance of the piezo. The voltage on the shaker piezo was ~ 70 V to produce a cavity detuning greater than 1 kHz. The waveform of the stimulus pulse on the shaker piezo mimics the RF pulse on the cavity with a rise time of 1.2 ms and the flat-top of 0.8 ms. The flat-top region is where the beam is accelerated. This region is where the piezo should minimize the detuning to reduce the overall RF power needed to accelerate the beam. Note that the time width of the flat-top can be changed, and the one used in this study is the same for various pulse linacs. The LFD during the flat-top is ~1.5 kHz. The results from the square wave pulse show that the cavity response during the flat-top is similar to the response in a real cavity at 2 K when pulse with an RF signal (see Fig. 3). The detuning during this time goes down linearly as expected. Note that even after the pulse is turned off, the cavity still experiences detuning.

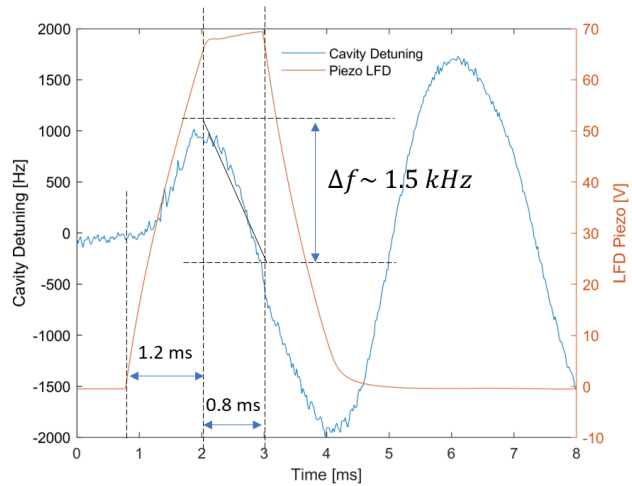

Figure 3: Simulated LFD pulse with a square wave pulse on the shaker piezo.

The goal of the LFD resonance control with the piezo is to minimize the cavity detuning during the flat-top region. This was achieved by using a sine wave with a frequency that will cause destructive interference with the cavity detuning response due to the LFD pulse. Three parameters need to be optimized for the sine wave: the frequency, the amplitude, and the time advance from the flat-top. These three parameters are optimized by trial and error in LabVIEW. The right frequency can be found by taking the FFT of the step response from Fig. 3, the result is shown in Fig 4. The most prominent frequencies excited are 175 Hz and 231 Hz. These frequencies were the first to be implemented for resonance control, but the results yielded poor frequency detuning compensation. Note that the sine wave amplitude and advance were optimized for these trials.

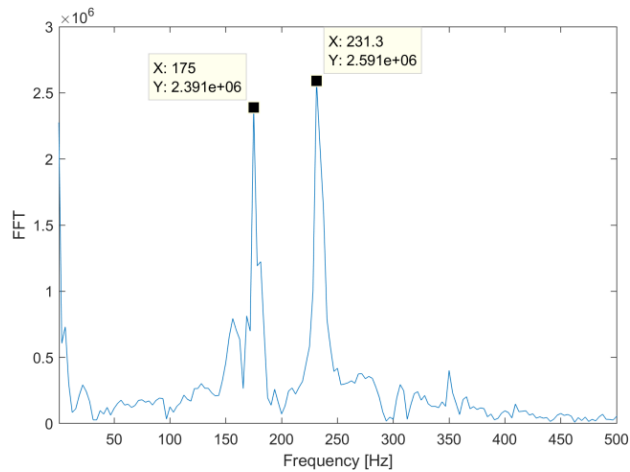

Figure 4: FFT of the cavity detuning from the simulated LFD pulse from the shaker piezo.

A sine wave of 160 Hz by trial and error resulted in a flat detuning line. The cavity response from the sine wave is shown in Fig. 5. There is a 1 ms delay between when the sine wave stimulus starts and when the cavity response is seen. This delay is due to the electronics and largely from the tuner system. The control piezo voltage is also smaller compared to the shaker piezo. As with the response from the shaker piezo in Fig. 3, the cavity detuning has a residual response from the sine pulse even after it is turned off.

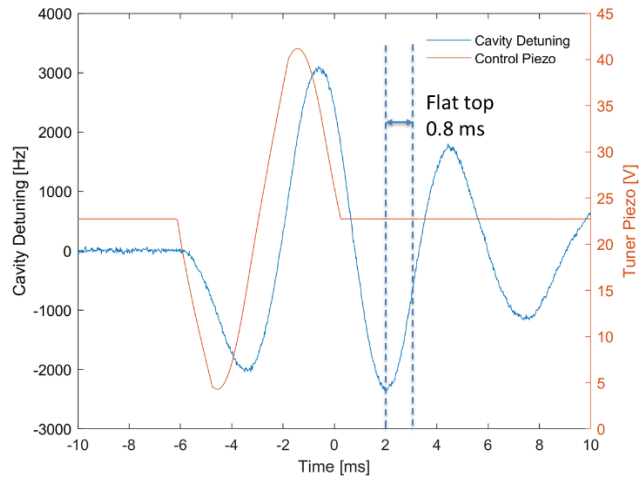

Figure 5: 160 Hz single cycle sine wave stimulus pulse applied to the piezo tuner (red). The cavity detuning response is shown in the blue line. Note the 1 ms delay from the control piezo signal to the cavity detuning.

## RESULTS

The single cycle sine wave applied to the control piezo advanced by 7 ms from the LFD pulse produced the best cavity detuning compensation (see Fig. 6). Figure 6 shows the cavity detuning, the piezo voltage on the shaker piezo, and control piezo voltage on the right axis. The advance between the control piezo and the LFD pulse was optimized by trial and error. The optimal stimulus pulse was a single sine wave with a voltage of 35 V peak-to-peak. The detuning at the flat-top dropped from 1.4 kHz to 140 Hz (see Fig. 7). Figure 7 is a close-up of the cavity detuning in the flat-top (during 0.8 ms). The detuning was thus reduced by a factor of 10. In future iterations, the piezo bias must also be optimized to set the cavity detuning line closer to 0 Hz.

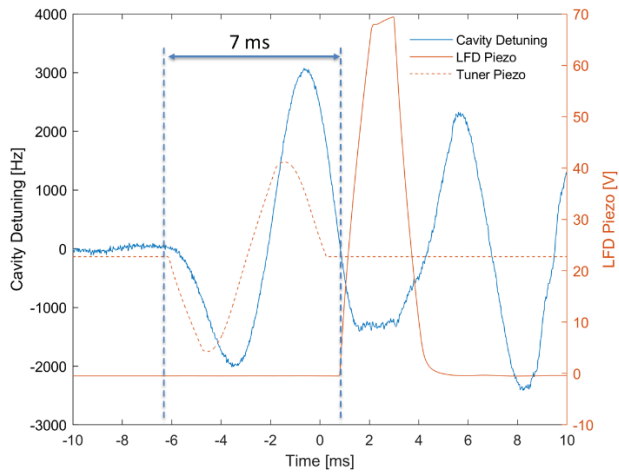

Figure 6: Control pulse which results in a flat cavity detuning. A sine wave of 160 Hz advanced by 7 ms from the LFD pulse was used. The left axis is for the cavity detuning, and the right axis is for the voltage of the piezos.

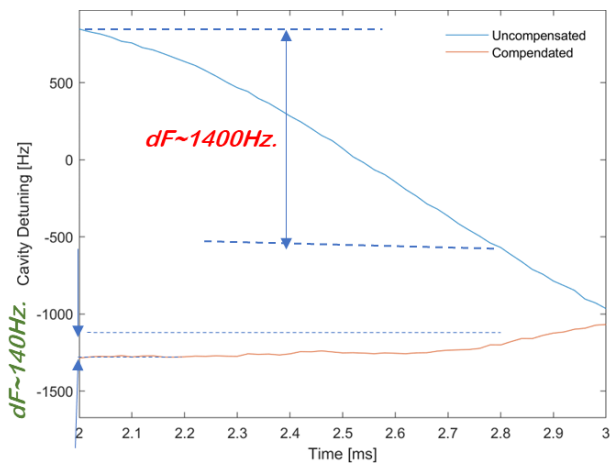

Figure 7: Comparison of cavity detuning with piezo control on and off. The results show that when the LFD pulse is compensated, the cavity detuning is decreased by a factor of 10.

## CONCLUSION

The study demonstrated that operation with warm SRF cavity, equipped with tuner, could be successfully used to develop LFD compensation algorithms. The LCLS-II double lever tuner was installed on a dressed 1.3 GHz cavity and used to demonstrate compensation of the dynamic LFD. The cavity vibration induced by the square wave pulse on the shaker piezo simulated the vibrations induced from LFD well. This simulated LFD generated a cavity detuning of about 1.4 kHz. A sine wave was applied to the control piezo to cause destructive interference with the resulting cavity detuning from the simulated LFD pulse. The results show that the detuning is decreased from 1.4 kHz to 140 Hz demonstrating that the LCLS-II tuner frame can compensate LFD. Further studies are planned to make the program automatic and apply a more advanced LFD compensation algorithms such as in [2].